# Efficiency of attack strategies on complex model and real-world networks

Michele Bellingeri[1*], Davide Cassi[1], Simone Vincenzi[2,3]

[1]Dipartimento di Fisica, Università di Parma, via G.P. Usberti, 7/a, 43124 Parma, Italy

[2]Center for Stock Assessment Research (CSTAR) and Department of Applied Mathematics and Statistics, University of California Santa Cruz, 110 Shaffer Road, 95060 Santa Cruz, CA, US.

[3]Dipartimento di Elettronica, Informazione e Bioingegneria, Politecnico di Milano, Via Ponzio 34/5, I-20133 Milan, Italy

* Corresponding author: michele.bellingeri@nemo.unipr.it

## Abstract

We investigated the efficiency of attack strategies to network nodes when targeting several complex model and real-world networks. We tested 5 attack strategies, 3 of which were introduced in this work for the first time, to attack 3 model (Erdos and Renyi, Barabasi and Albert preferential attachment network, and scale-free network configuration models) and 3 real networks (Gnutella peer-to-peer network, email network of the University of Rovira i Virgili, and immunoglobulin interaction network). Nodes were removed sequentially according to the importance criterion defined by the attack strategy. We used the size of the largest connected component (*LCC*) as a measure of network damage. We found that the efficiency of attack strategies (fraction of nodes to be deleted for a given reduction of *LCC* size) depends on the topology of the network, although attacks based on the number of connections of a node and betweenness centrality were often the most efficient strategies. Sequential deletion of nodes in decreasing order of betweenness centrality was the most efficient attack strategy when targeting real-world networks. In particular for networks with power-law degree distribution, we observed that most efficient strategy change during the sequential removal of nodes.

## 1. Introduction

The resilience of real-world complex networks, such as Internet, electrical power grids, airline routes, ecological and biological networks [1,2,3,4,5,6] to "node failure" (i.e. node malfunctioning or removal) is a topic of fundamental importance for both theoretical and applied network science. Node failure can cause the fragmentation of the network, which has consequences in terms of system performance, properties, and architecture, such as transportation properties, information delivery efficiency and the reachability of network components (i.e. ability to go from node of the network to another) [3].

Several studies [3,7,8,9] have investigated the resilience of model networks using a number of "attack strategies", i.e. a sequence of node removal according to certain properties of the nodes [2,3,7]. A widely-applied attack strategy consists in first ranking the nodes with respect to an importance criterion (e.g. number of connections or some measures of centrality) and then remove the nodes sequentially from the most to the least important according to the chosen criterion until the network either becomes disconnected or loses some essential qualities [3,10]. However, little is known on how the efficiency (i.e. fraction of nodes to be deleted for a given change in the network) of attack strategies varies when considering differing real-world and model networks.

In addition, an interesting - although underappreciated - issue is how the relative efficiency of attack strategies may change during the attack. For example, an attack strategy might be more efficient when the targeted (i.e. under attack) network is still pristine, while other strategies might be more efficient when the network has already been fragmented and some of its properties been compromised. Testing the efficiency of the different attack strategies when targeting different networks may also allow to identify the most important nodes for network functioning, and therefore which nodes should be primarily protected, as in the case of computer [11] or ecological networks [6,12,13,14], or removed, as in the case of immunization/disease networks [15].

In this work, we test the efficiency of both well-known attack strategies and new strategies introduced for the first time in this paper when targeting either model or real-world networks, using the size of the largest connected component (*LCC*) (i.e. the largest number of nodes connected among them in the network, [2]) as a measure of network damage. We found for model networks that the best strategy to reduce the size of the *LCC* depended on the topology of the network that is attacked. For real-world networks, the removal of nodes using betweenness centrality as importance criterion was consistently the most efficient attack strategy. In addition, we found that for some networks an attack strategy can be more efficient than others up to a certain fraction of nodes removed, but other attack strategies can become more efficient after that fraction of nodes has been removed.

## 2.Methods

### 2.1 Attack strategies

We attacked the networks by sequentially removing nodes following some importance criteria. We compared the efficiency of a pool of attacks strategies, some of which have been already described in the literature while others are introduced in this work for the first time. Most of the analyses on the robustness of network when have investigated the effect of removing nodes according to their rank (i.e. number of links) or some measures of centrality [3,10,16]. In this work, we introduce new attack strategies that focus entirely or in part on less local properties of a node, in particular its number of second neighbors, as explained in detail below.

Several indexes and measures have been introduced to describe network damage. In this work, we use the size of the largest connected component (*LCC*), i.e. the size of the largest connected sub-graph in the network [2,3], as a measure of network damage during the attack. A faster decrease in the size of the *LCC* indicates a more efficient attack strategy. In order to compare attack strategies across networks, we used the normalized *LCC* size with respect to the starting *LCC* size, i.e. the number of nodes in the *LCC* before any removal.

For each attack strategy, we applied both the recalculated and non-recalculated methods. In the recalculated method, the property of the node relevant for the attack strategy (e.g. number of links) was recalculated after each node removal. In the non-recalculated method, the property of the node was computed before the first node removal and was not updated during the sequential deletion of nodes. An attack strategy is less efficient than another when a higher the fraction of nodes has to be removed to reduce the *LCC* to zero (or any other size). With $q$ we indicate the fraction of nodes removed during the sequential removal of nodes.

We used 2 attack strategies that have been already described in the literature. *First-degree neighbors (First)*: nodes are sequentially removed according to the number of first neighbors of each node (i.e. node rank). In the case of ties (i.e. nodes with the same rank), the sequence of removal of nodes is randomly chosen. *Nodes betweenness centrality (Bet)*: nodes are sequentially removed according to their betweenness centrality, which is the number of shortest paths from all vertices to all others that pass through that node [3,17].

We introduced in this paper the following new attack strategies. *Second-degree neighbors (Sec)*: nodes are sequentially removed according to the number of second neighbors of each node. Second neighbors of node $j$ are nodes that have a node in common with, but are not directly connected to, node *j*. *First + Second neighbors (F+S)*: nodes are deleted according to the sum of first and second neighbors of each node. *Combined first and second degree* (*Comb*): nodes are removed according to their rank. In the case of ties, nodes are removed according to their second degree.

For all attack nodes were sequentially removed from most to least connected, or in case of *Bet*, from higher to lower betweenness centrality.

## 2.2 Networks

We tested the attack strategies described in Section 2.2 on (*i*) 3 types of model networks and (*ii*) 3 real world networks.

The networks we used are undirected and unweighted graphs in which nodes are connected by links or edges, and rank *k* of a node is the number of links of that node. Each link may represent several real world interactions. For instance, in social networks links between nodes represent interactions between individuals or groups, such as co-authorship in scientific publications or friendship [2]. In cellular networks, nodes are chemicals species connected by chemical reactions [18], while in ecological networks links describe the trophic interactions between species or group of species, e.g. the energy and matter passing from prey to predator [6,14,19,20].

### 2.2.1 Model networks

We tested the attack strategies on (i) Erdos and Renyi graphs [21], (ii) Barabasi and Albert preferential attachment networks [2], and (iii) scale-free network configuration models [22]. For each model network, we tested models of different size, as explained below. Since each model network is a random realization of the network-generating mechanism, we tested the attack strategies on 50 random realizations of each model network used the mean of the normalized *LCC* size at each fraction $q$ of nodes removed as a measure of network damage. We observed a small variation of *LCC* size at each fraction $q$ of nodes removed across different realizations of networks, thus the mean *LCC* size across replicates well represents the overall behavior of the attack strategy.

The Erdos and Renyi (*ER*) model generates a random graph with $N$ nodes connected by $L$ links, which are chosen randomly with an occupation probability $p$ from $L_{max} = N(N-1)/2$ possible links, i.e. $p$ is the proportion of realized links from $L_{max}$. The expected number of links is $<L> = (N^2p)/2$ and the expected rank of a node is $<k> = Np$. The random graph can be defined by the number of nodes $N$ and the probability $p$, i.e. $ER(N,p)$ [21]. We analyzed *ER* graphs with different values of $N$ and $p$, namely: $ER(N = 500, p = 0.008)$, $ER(1\,000, 0.004)$, $ER(10\,000, 0.0004)$.

The Barabasi and Albert preferential attachment network (*BA*) is created starting from few isolated nodes and then growing the network by adding new nodes and links [2]. At each step in the creation of the network, one node and $m$ outgoing links from the new node are added to the network. The probability $\theta$ that the new node will be connected to node $i$ already in the network is function of the degree $k_i$ of node $i$, such that $\theta(k_i) = k_i / \sum_{j=N}^{j=1} k_j$ (i.e. preferential attachment, since more connected nodes are more likely to be connected to the new node) [2]. The *BA* network is defined by parameters $N$ and $m$, i.e. $BA(N,m)$. We built *BA* scale free networks with parameters $BA(N=500, m = 2)$, $BA(1\,000, 2)$, $BA(10\,000, 2)$.

We created networks with power-law degree distribution using the configuration model for generalized random graphs [2,22,]. This model is defined as follows. A discrete degree distribution $P(K = k) = k^{-\alpha}$ is defined, such that $P(k)$ is the proportion of nodes in the network having degree $k$. The maximum node degree $k_{\max}$ is equal to $N$, where $N$ is the number of nodes. The domain of the discrete function $P(k)$ becomes (1, $k_{max}$). We generated the degree sequence of the nodes by randomly drawing $N$ values $k_1,\ldots,k_n$ from the degree distribution. Then, for each node $i$ we drew a link with node $j$ with probability $P(k_i)P(k_j.)$ A scale free configuration model network is defined by parameters $N$, $\alpha$ and $<k>$. We analyzed scale free network with parameters $CM(N = 500, \alpha = 2.5, <k> = 3.8)$, $CM$(1 000,2.5,3.8), $CM$(10 000,2.5,3.9).

#### 2.2.2 Real world networks

We tested the attack strategies on the following real-world networks: (*i*) The Gnutella P2P (peer-to-peer) network (*Gnutella*) [24], (*ii*) the email network of the University Rovira i Virgili (URV) in Tarragona, Spain (*Email*) [25], and (*iii*) the immunoglobulin interaction network (*Immuno*) [26]. Nodes of *Gnutella* ($N$=8846, $L$=31839) represent hosts in the peer-to-peer network while links represent connections between the hosts [24]. *E-mail* ($N$=1134, $L$=10902) provides a representative example of the flow of information within a human organization [25]. *Immuno* is the undirected and connected graph of interactions in the immunoglobulin protein ($N$ = 1316, $L$ = 6300) where nodes represent amino acids and two amino acids are linked if they interact in the immunoglobulin protein [26].

## 3. Results

### 3.1 Non-Recalculated method

#### 3.1.1 Model networks (Fig. 1)

***ER***: For all sizes of networks, the 5 attack strategies were equally efficient in reducing the size of the *LCC* up to $q$ ~= 0.2. For $q$ >0.2, *First* was the most efficient strategy to reduce the size of the *LCC*.

***CM***: For $N$ = 500, *Comb* was the most efficient strategy early in the removal sequence., while *First* became the most efficient strategy after $q$ > 0.1. For $N$ = 1 000, *Comb*, *Bet*, and *First* had the same efficiency. For $N$ = 10 000, *Comb*, *Bet*, and *First* were equally efficient up to $q$ = 0.1, while for $q$ > 0.1 *First* was the most efficient strategy.

***BA***:  For $N$ = 500, *First*, *Comb* and *Bet* were equally efficient in reducing the size of the *LCC*. For bigger networks, *First*, *Comb* and *Bet* were equally efficient up to $q$ = 0.8 ($N$ = 1 000) and $q$ = 0.5 ($N$ = 10 000). Then, *Bet* became more efficient than *First* and *Comb.*

### 3.1.2 Real-world networks (Fig. 2)

***Email***: *Bet* was the most efficient strategy to reduce *LCC* up to $q$~= 0.3. For greater fractions of nodes removed, *First* and *Comb* were slightly more efficient than *Bet*.

***Immuno***: *Bet* was distinctly more efficient than other strategies up to $q$ = 0.55. For $q$ > 0.55, all strategies were equally efficient.

***Gnutella***: *Bet* was the most efficient attack strategy.

## 3.2 Recalculated

### 3.2.1 Model networks (Fig. 3)

**ER**: *First* and *Comb* were the most efficient strategies to reduce the *LCC* up to $q$~=0.2. Beyond that fraction of nodes removed, *Bet* became more efficient than *First. Sec* was the least efficient strategy.

**CM**: *Comb* was the most efficient strategy up to *q*~=0.1. Beyond that, *Bet* was the most efficient strategy, while *Sec* was the least efficient strategies.

**BA**: *Comb* was the most efficient strategy up to *q*~=0.1. *First*, *F+S* and *Bet* attack induced a slightly slower decrease in *LCC* size. When *q* reached 0.1, *Bet* became the most efficient strategy. *Sec* was the least efficient strategy.

#### 3.2.2 Real-world networks (Fig. 4)

***Email***: All attack strategies were equally efficient up to $q = 0.12$. For $q > 0.12$ *Bet* was the most efficient attack strategy.

***Immuno***: *Bet* was largely the most efficient attack strategy.

***Gnutella***: All attack strategies were equally efficient up to $q = 0.1$. For $q > 0.1$ *Bet* was the most efficient attack strategy.

## 4. Discussion

We discuss the following main results of our work: (*i*) attacks were largely more efficient (i.e. smaller fraction of nodes deleted to achieve the same reduction in *LCC* size) with the recalculated than with the non-recalculated method; (*ii*) the efficiency of attack strategies on model networks depended on network topology; (*iii*) the sequential removal of nodes according to their betweenness centrality was the most efficient attack to real-world networks; (iv) for some networks, the relative efficiency of attack strategies changed during the removal sequence.

We found that the recalculated method provided more efficient attacks than the non-recalculated method, i.e. for a given fraction of nodes removed a larger reduction of *LCC* was obtained with the recalculated method. This result confirms the findings of other analyses on robustness of networks [2,3], which found that updated information on the topology of the system after each removal allowed for more efficient attacks to networks.

However, non-recalculated attack strategies are implemented in various relevant settings and are equivalent in practice to the simultaneous removal of nodes, as it happens in the case of vaccination campaigns (i.e. vaccinating at the same time nodes of the contact network with the highest probability of acquiring or transmitting the disease) or attacks to computer networks [11].

For model networks, the efficiency of the attack strategies depended on network topology. In the case of networks with power-law degree distribution, the efficiency of the attack strategies depended also on network size. Across all model networks and considering both the non-recalculated and recalculated methods, attack strategies based on either the betweenness centrality of nodes or node rank were the most efficient ones. However, the sequential deletion of nodes according to their betweenness centrality was consistently the most efficient attack strategy to real-world networks, with the only exception of the attack to the *Email* network with the non-recalculated method. While in some cases *Bet* was slightly more efficient than other strategies in reducing the size of the largest connected component, in others *Bet* was largely the most efficient strategy. For example, in the immunoglobulin interaction network, deleting a very small fraction of nodes with high betweenness centrality reduced the size of the normalized *LCC* of more than 60% using either the recalculated and non-recalculated method, while for the same fractions of nodes removed other attack strategies were able to obtain only a 1-5% reduction in *LCC* size. Betweenness centrality describes how "central" a node is in the network by considering the fraction of shortest paths that pass through that node [17]. Nodes with betweenness centrality greater than 0 play a major role in connecting areas of the network that would otherwise be either sparsely connected or disconnected [23]. This makes betweenness centrality an important centrality measure for a social, technological, computer, and biological networks. The higher efficiency of the strategy based on node betweenness centrality with respect to the attack based on node rank in real-world networks can be explained by the fact that in real-world networks some of the critical nodes (i.e. nodes whose persistence strongly contribute to maintaining network integrity) are either not highly linked, or that the highly linked nodes are not located in the network core [23].

The newly-introduced *Combined* attack strategy, when recalculated, was the most efficient strategy to decrease *LCC* size in the scale free network configuration model and in the Barabasi-Albert model up to $q$ = 0.1, performing slightly better than the attack based on node rank (*First* strategy). The *Combined* attack first select nodes according to their rank, then, in the case of ties (i.e. nodes with the same rank), it sequentially removes nodes according to their second degree. On the contrary, in the case of ties *First* randomly chooses the removal sequence for the nodes with the same rank. Thus, at the beginning of the attack to the network, when two or more major hubs have the same number of links to other nodes, selecting to remove first the hub with the greatest second degree causes a faster decrease in *LCC* size than to randomly select the removal sequence for those hubs.

Later in the attack, the *Combined* strategy was less efficient than the *First* strategy to attack scale free networks; this might be due to the fact that after a certain fraction of hubs has been deleted, removing first (in the case of ties) the node(s) with the highest second degree(s) would remove more peripheral and less important nodes. Those nodes are less likely to be part of the largest connected component.

Lastly, the efficiency of attack strategies changed along the sequential removal of nodes. This was particularly evident for networks with power-law degree distribution. It follows that the percolation threshold, i.e. the fraction of nodes removed for which the size of the largest connected component reaches zero, might be for some networks little correlated with the fraction of nodes to be removed in order to reduce the largest connected component to a size greater than 0. This result has important implications for applied network science and deserves further investigations. For example, in the case of immunization strategies, choosing the attack strategy according to the percolation threshold may be of little use when the goal is to reduce as much as possible the size of *LCC* with just a few targeted immunizations.

## Acknowledgements

We thank Riccardo Campari, Daniele Marmiroli and Prof. Alessio Camobreco for useful comments on a previous version of the manuscript.

# Figures

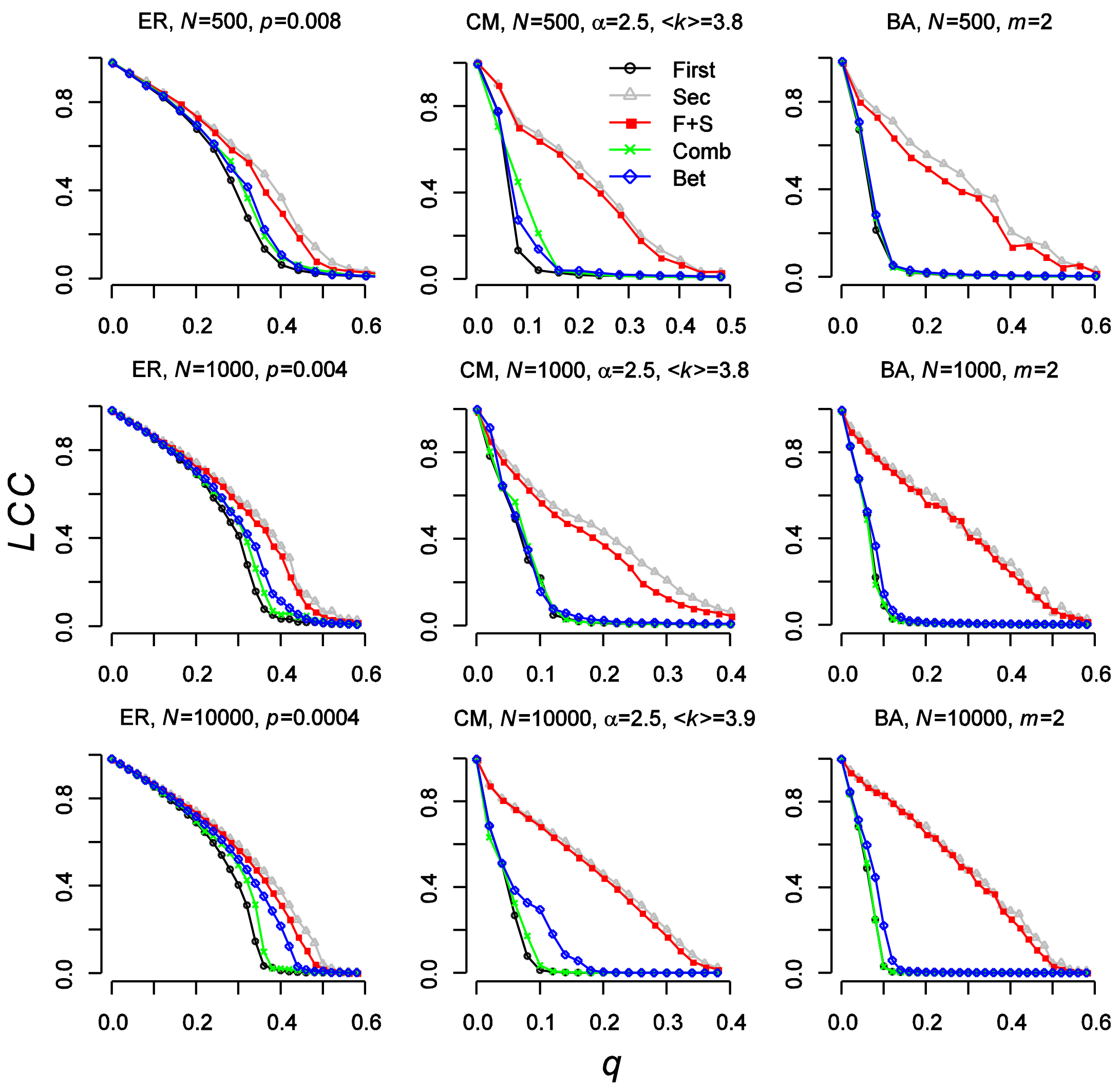

**Figure 1**. Size of normalized *LCC* and the fraction *q* of nodes removed for non-recalculated targeted attacks to model networks. Points are plotted every 20 nodes removed for networks with *N* = 500 and *N* =1 000, and every 200 nodes removed for *N* =10000.

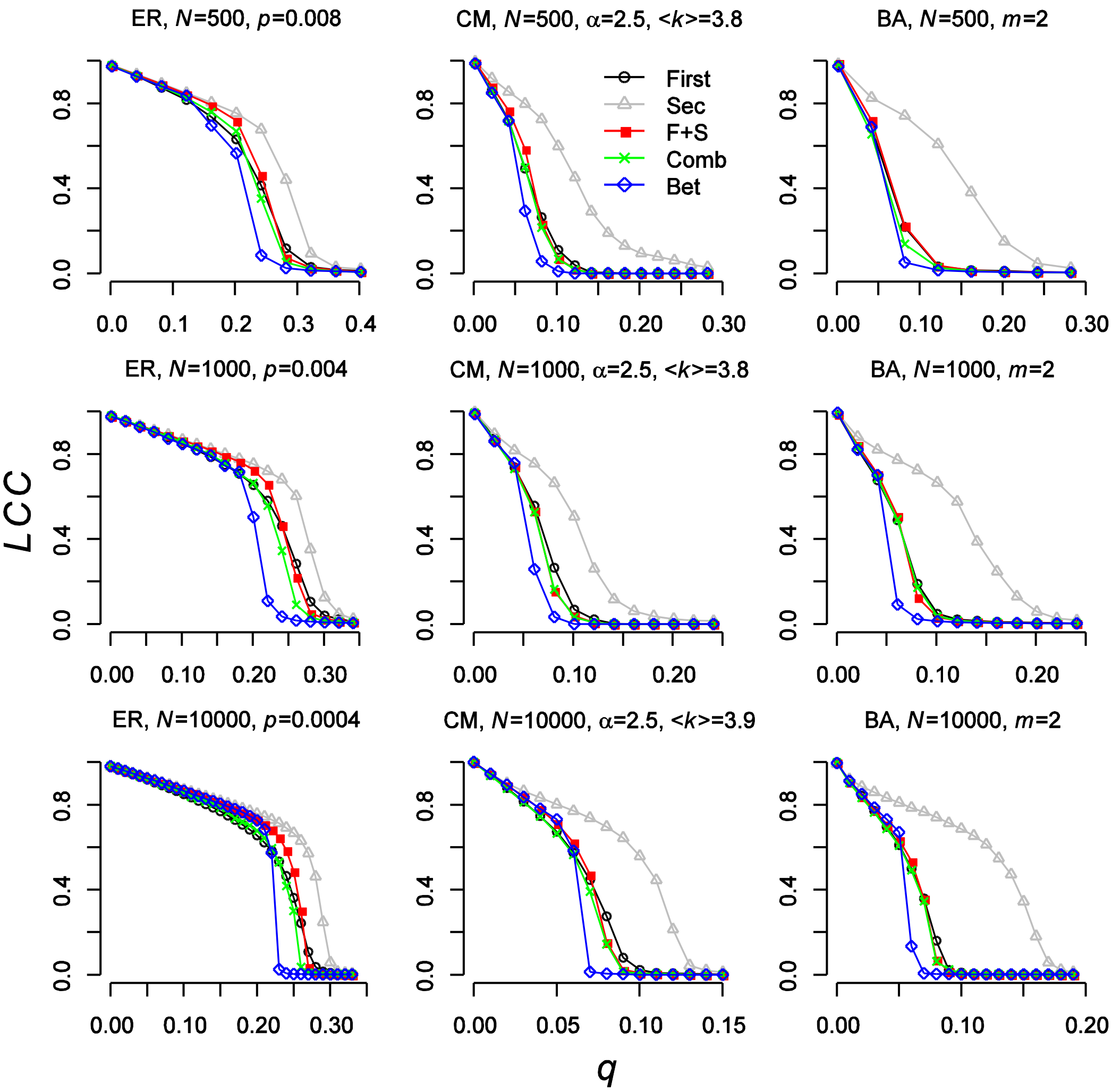

**Figure 2**. Size of normalized *LCC* and the fraction *q* of nodes removed for non-recalculated targeted attacks to real-world networks. Points are plotted every 50 nodes removed for *Email* and *Immuno* networks, and every 200 nodes removed for *Gnutella*.

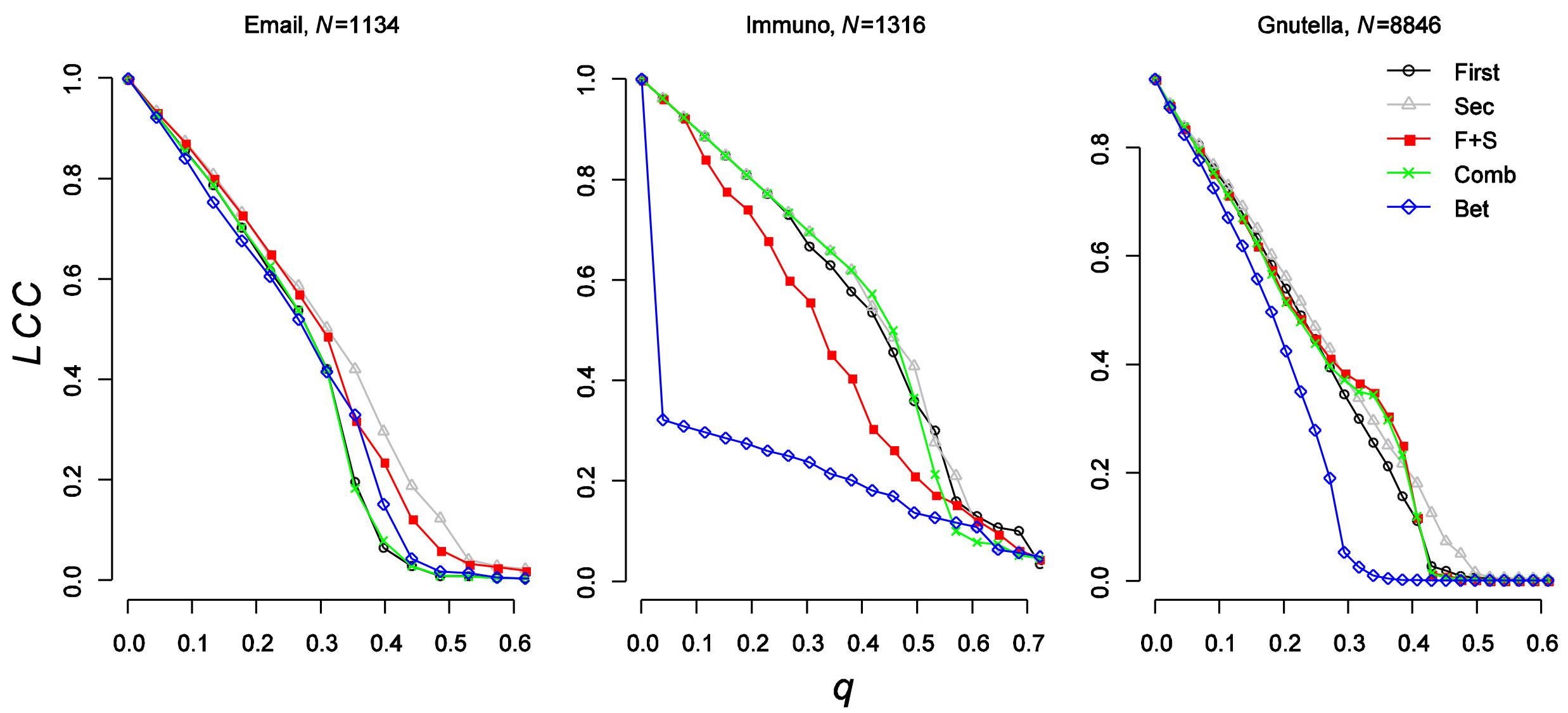

**Figure 3**. Size of normalized *LCC* and the fraction *q* of nodes removed for recalculated targeted attacks to model networks. Points are plotted every 20 nodes removed for networks with $N = 500$ and $N$ =1 000, and every 200 nodes removed for $N$ =10000.

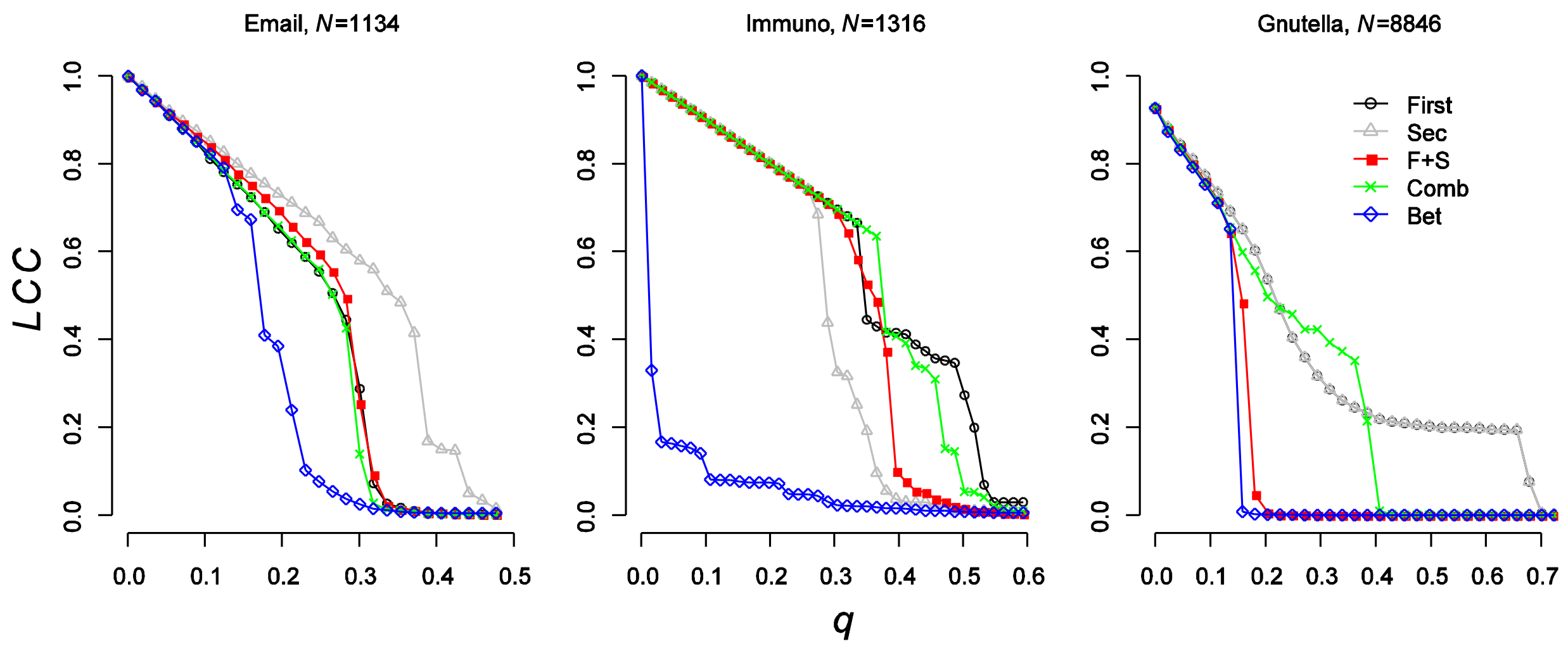

**Figure 4**. Size of normalized *LCC* and the fraction *q* of nodes removed for recalculated targeted attacks to real-world networks. Points are plotted every 50 nodes removed for *Email* and *Immuno* networks, and every 200 nodes removed for *Gnutella*.